\documentclass[submission, Phys]{SciPost}

\usepackage{amsmath}
\usepackage[english]{babel}
\usepackage{graphicx}
\usepackage[charter]{mathdesign}
\usepackage{bm}
\usepackage{subcaption}
\usepackage{array}

\newcolumntype{C}{>{$\displaystyle}c<{$}}
\newcolumntype{L}{>{$\displaystyle}l<{$}}
\newcolumntype{R}{>{$\displaystyle}r<{$}}

\renewcommand\a{\alpha}
\renewcommand\b{\beta}

\newcommand\av{\mathbf{a}}
\newcommand\bv{\mathbf{b}}

\newcommand\dn{\downarrow}

\newcommand\Ev{\mathbf{E}}

\newcommand\Gv{\mathbf{G}}

\newcommand\Kv{\mathbf{K}}

\newcommand\kvt{\mathbf{\tilde k}}

\newcommand\om{\omega}

\newcommand\qv{\mathbf{q}}

\newcommand\Rv{\mathbf{R}}
\newcommand\rv{\mathbf{r}}
\newcommand\s{\sigma}
\newcommand\Sigmav{\bm{\Sigma}}

\DeclareMathOperator\Tr{Tr}

\newcommand\tv{\mathbf{t}}
\newcommand\unit{\mathbf{1}}
\newcommand\up{\uparrow}
\newcommand\Vv{\mathbf{V}}

\begin{document}

\begin{center}{\Large \textbf{
Charge order and antiferromagnetism in twisted bilayer graphene from the variational cluster approximation 
}}\end{center}

\begin{center}
B. Pahlevanzadeh\textsuperscript{1,2},
P. Sahebsara\textsuperscript{1},
D. S\'en\'echal\textsuperscript{2*}
\end{center}

\begin{center}
{\bf 1} Department of Physics, Isfahan University of Technology, Isfahan, Iran
\\
{\bf 2} D\'epartement de physique and Institut quantique, Universit\'e de Sherbrooke, Sherbrooke, Qu\'ebec, Canada J1K 2R1
\\
* david.senechal@usherbrooke.ca
\end{center}

\begin{center}
\today
\end{center}


\section*{Abstract}
{\bf
We study the possibility of charge order at quarter filling and antiferromagnetism at half-filling in a tight-binding model of magic angle twisted bilayer graphene.
We build on the model proposed by Kang and Vafek~\cite{kang2018}, relevant to a twist angle of $1.30^\circ$, and add on-site and extended density-density interactions.
Applying the variational cluster approximation with an exact-diagonalization impurity solver, we find that the system is indeed a correlated (Mott) insulator at fillings $\frac14$, $\frac12$ and $\frac34$.
At quarter filling, we check that the most probable charge orders do not arise, for all values of the interaction tested. At half-filling, antiferromagnetism only arises if the local repulsion $U$ is sufficiently large compared to the extended interactions, beyond what is expected from the simplest model of extended interactions.
}

\vspace{10pt}
\noindent\rule{\textwidth}{1pt}
\tableofcontents\thispagestyle{fancy}
\noindent\rule{\textwidth}{1pt}
\vspace{10pt}

\section{Introduction}

The observation of correlated insulators and superconductivity in twisted bilayer graphene (TBG) \cite{cao2018, cao2018a} has
inaugurated the new field of twistronics.
This discovery was motivated by the prediction that, for a few small ``magic'' twist angles, the band structure of a twisted graphene bilayer would contain a low-energy manifold of flat bands, well separated from the other bands and forming a strongly correlated electronic subsystem.~\cite{bistritzer2011,suarezmorell2010,tramblydelaissardiere2016}.
So far the superconducting order parameter symmetry of TBG is not known, although there are numerous predictions.
The precise nature of the insulating state (pure Mott insulator or broken symmetry phase) is not precisely known either.
The goal of this paper is to analyse the insulating state of TBG at quarter- and half-filling and to ascertain whether it is a pure Mott state or a broken symmetry state, either a charge-density wave (quarter filling) or an antiferromagnet (half-filling).
We will conclude that it is indeed a pure Mott state.

This paper is an extension of our previous work \cite{pahlevanzadeh2021} on the superconducting state of TBG.
We will use the same premise: We will start from the tight-binding model proposed by Kang and Vafek~\cite{kang2018}, based on the microscopic analysis of Moon and Koshino~\cite{moon2012}. 
However, instead of applying cluster dynamical mean field theory (CDMFT) as in Ref.~\cite{pahlevanzadeh2021}, we will apply another cluster method, the variational cluster approximation (VCA), based on a 12-site cluster.
In addition, we will include extended interactions, which were neglected in Ref.~\cite{pahlevanzadeh2021} and will extend the VCA by a mean-field treatment of inter-cluster interactions. 
Since the model studied is nearly particle-hole symmetric, the conclusions reached at quarter filling also apply at three-quarter filling.

\section{The low-energy model}\label{sec:model}

\begin{figure}[ht]\centering
\includegraphics[width=0.5\hsize]{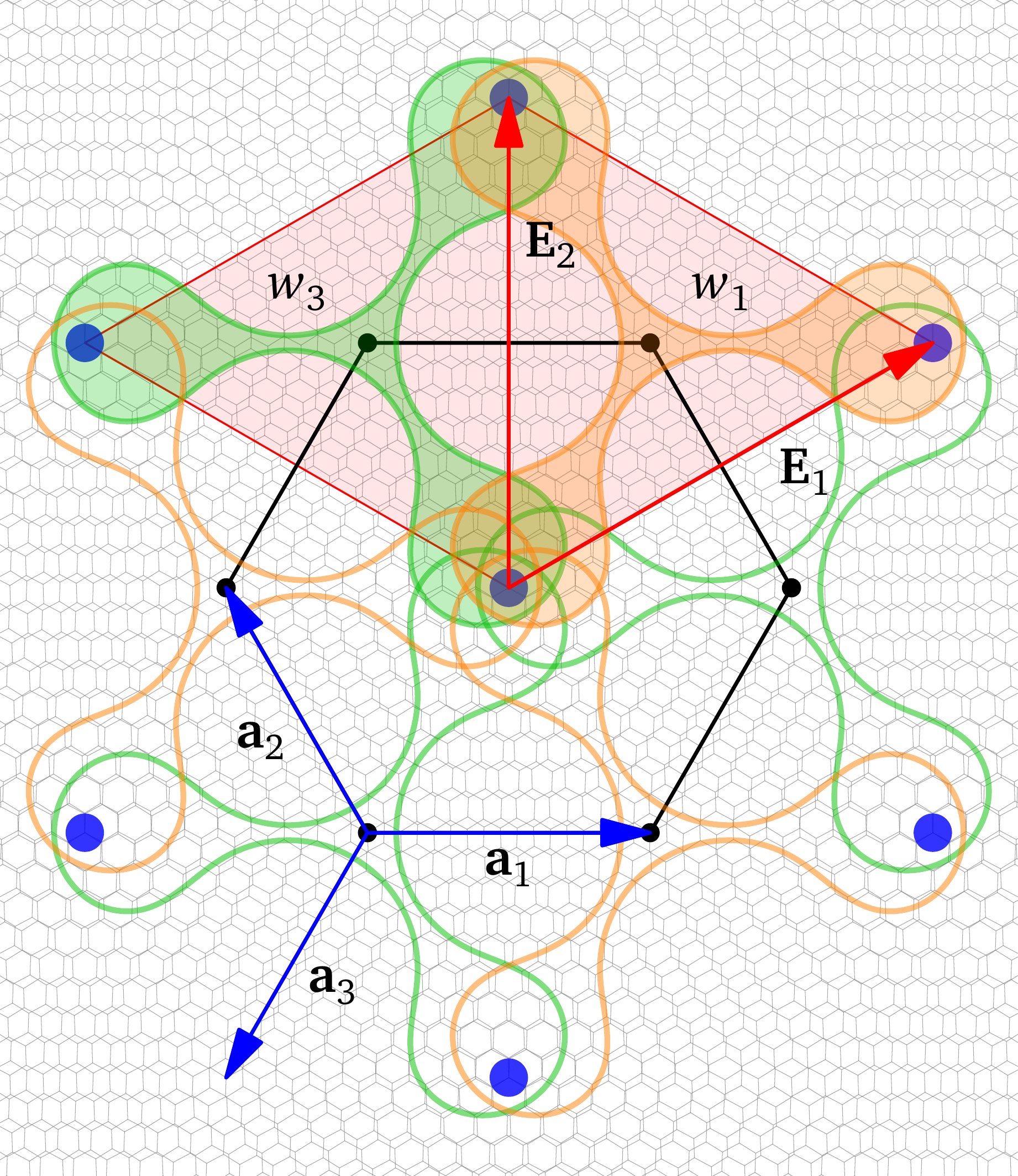}
\caption{Schematic representation of the Wannier functions $w_1=w_2^*$ (orange) and $w_3=w_4^*$ (green) on which our model Hamiltonian is built.
The charge is maximal at the AA superposition points (blue circles) forming a triangular lattice.
The Wannier functions are centered on the triangular plaquettes that form a graphene-like lattice (black dots), whose unit cell is shaded in red.
The basis vectors $\Ev_{1,2}$ of the moir\'e lattice are shown (they are also basis vectors of the graphene-like lattice of Wannier functions), as well as the elementary nearest-neighbor vectors $\av_{1,2,3}$. This figure is borrowed from Ref.~\cite{pahlevanzadeh2021}.
}
\label{fig:Wannier}
\end{figure}

Among the various tight-binding Hamiltonian proposed for the low-energy bands of TBG~\cite{angeli2018, moon2012, kang2018, yuan2018a}, we adopt the one described in Ref.~\cite{kang2018}. 
This model features four Wannier orbitals per unit cell (labeled $w_{1,2,3,4}$), with maximal symmetry, on an effective honeycomb lattice, appropriate for a twist angle $\theta=1.30^\circ$.
Each site of the honeycomb lattice is associated with two Wannier orbitals, which it is convenient to imagine located on two different layers, containing respectively the orbitals $w_{1,4}$ and the orbitals $w_{2,3}$.
The Wannier orbitals of one layer are schematically illustrated on Fig.~\ref{fig:Wannier}, borrowed from Ref.~\cite{pahlevanzadeh2021}.
We will only retain the largest hopping integrals among those computed in Ref.~\cite{kang2018}; see Table~\ref{table:hopping} (the notation used is that of Ref.~\cite{kang2018}).
The most important hopping terms are between Wannier orbitals $w_1$ and $w_4$ and between $w_2$ and $w_3$, i.e., between graphene sublattices, within a given layer. The inter-layer hopping terms are much smaller, the largest of which being $t_{13}[0,0]$.

\definecolor{Green}{rgb}{0,0.7,0}
\begin{table}[ht]\centering
	$\vcenter{\hbox{\begin{tabular}{LL}
			\hline\hline
			\mathrm{symbol} & \mbox{value (meV)} \\ \hline
			{\color{white}\bullet}~t_{13}[0,0] = \omega t_{13}[1,-1] = \omega^* t_{13}[1,0] & -0.011 \\
			{\color{red}\bullet}~t_{14}[0,0] = t_{14}[1,0] = t_{14}[1,-1] & \phantom{-}0.0177 + 0.291i \\
			{\color{blue}\bullet}~t_{14}[2,-1] = t_{14}[0,1] = t_{14}[0,-1] &  -0.1141 - 0.3479i \\
			{\color{Green}\bullet}~t_{14}[-1,0] = t_{14}[-1,1] = t_{14}[1,-2] & \\
			~~~= t_{14}[1,1] = t_{14}[2,-2] = t_{14}[2,0] & \phantom{-}0.0464 - 0.0831i \\ \hline\hline
			\end{tabular}}}$~~
	$\vcenter{\hbox{\includegraphics[width=4cm]{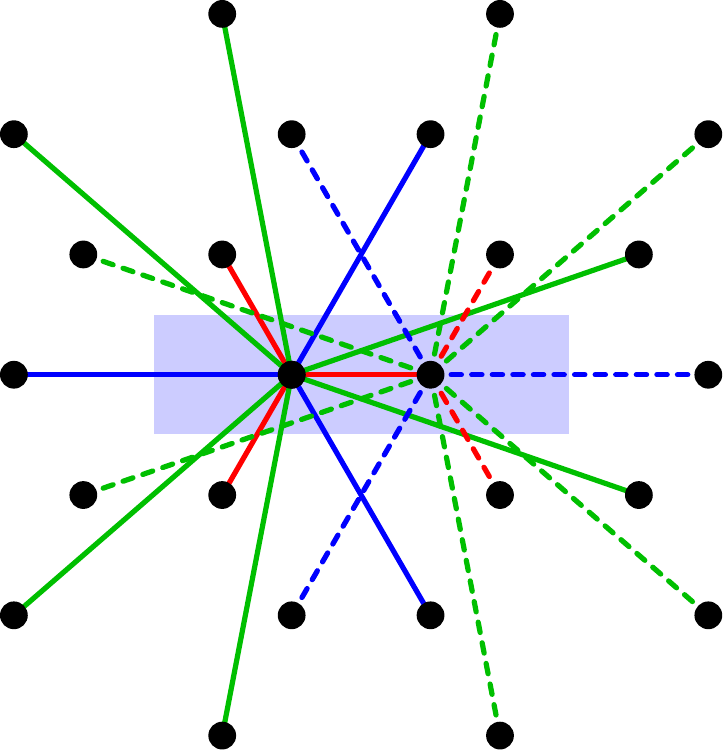}}}$
	\caption{Hopping amplitudes used in this work. They are the most important amplitudes computed in Ref.~\cite{kang2018}.
		Here $\om=e^{2\pi i/3}$ and the vector $[a,b]$ following the symbol represents the bond vectors in the $(\Ev_1,\Ev_2)$ basis shown on Fig.~\ref{fig:Wannier}.
		Note that $t_{23}=t_{14}^*$ and $t_{24}=t_{13}^*$.
		On the right: schematic view of the hopping terms $t_{14}$ within a given layer (the unit cell is the blue shaded area). Lines 2, 3, and 4 of the table correspond to the red, blue and green links, respectively. Dashed and full lines are for $t_{14}$ and $t_{23}$, respectively.
		\label{table:hopping}}
\end{table}

We now proceed to describe a simple model for interactions, derived from an on-site Coulomb repulsion at the AA sites~\cite{dodaro2018, xu2018c}:
\begin{equation}\label{eq:Hint}
H_{\rm int} = u\sum_{\Rv\in \mathrm{AA}} n_\Rv^2~~,
\end{equation}
where the sum is carried over AA sites and $n_\Rv$ is the total charge located at that site, to which contribute 12 Wannier orbitals (6 per layer).
Specifically, we could write
\begin{equation}\label{eq:nR}
n_\Rv = \frac13\sum_{i=1}^3 \left(n^{(1)}_{\Rv+\av_i}+n^{(1)}_{\Rv-\av_i}+n^{(2)}_{\Rv+\av_i}+n^{(2)}_{\Rv-\av_i}\right)
\end{equation}
where $n^{(\ell)}_\rv$ is the electron number associated with the Wannier orbital centered at the (honeycomb) lattice site $\rv$ on layer $\ell$.
The vectors $\pm\av_i$, indicated on Fig.~\ref{fig:Wannier}, go from each AA site to the six neighboring honeycomb lattice sites.
The factor of $\frac13$ above comes from the fact that each Wannier orbital has three lobes, i.e., is split across three AA sites.

Expressed in terms of the Wannier electron densities $n_\rv^{\ell}$, the interaction takes the form
\begin{equation}
H_{\rm int} = 
\frac12\sum_{\rv,\rv',\ell,\ell'} V_{\rv,\rv'}^{\ell,\ell'} n_{\rv}^\ell n_{\rv'}^{\ell'}
\end{equation}
where the factor of $\frac12$ avoids double counting when performing independent sums over sites and orbitals.
The Hubbard on-site, intra-orbital interaction $U$ is equal to $V_{\rv,\rv}^{\ell,\ell}$, since
\begin{equation}
V_{\rv,\rv}^{\ell,\ell}  n_{\rv\uparrow}^\ell n_{\rv\downarrow}^\ell = 
\frac12 V_{\rv,\rv}^{\ell,\ell} (n_{\rv\uparrow}^\ell+n_{\rv\downarrow}^\ell)(n_{\rv\uparrow}^\ell+n_{\rv\downarrow}^\ell) - \frac12 V_{\rv,\rv}^{\ell,\ell} n_\rv^\ell  \qquad(n_{\rv\sigma}^2=n_{\rv\sigma})
\end{equation}
Including on-site interactions in this form entails a compensation term $U/2$ to the chemical potential.

Careful counting from Eqs~(\ref{eq:Hint},\ref{eq:nR}) shows that
\begin{equation}\begin{aligned}\label{eq:values}
U &= \frac23 u &&\qquad \mbox{(on-site)}\\
V_{\rv\rv}^{(1,2)} &\equiv V_0 = \frac23 u = U &&\qquad \mbox{(same site, different layers)}\\
V_{\rv\rv'}^{(\ell,\ell')} &\equiv V_1 = \frac49 u = \frac23U &&\qquad \mbox{(1st neighbors)} \\
V_{\rv\rv'}^{(\ell,\ell')} &\equiv V_2 = \frac29 u = \frac13U &&\qquad \mbox{(2nd neighbors)} \\
V_{\rv\rv'}^{(\ell,\ell')} &\equiv V_3 = \frac29 u = \frac13U &&\qquad \mbox{(3rd neighbors)} 
\end{aligned}
\end{equation}
There are no interactions beyond third neighbors coming from a single AA site.
We will study this model by assuming the above relations between extended interactions $V_{0,1,2,3}$ and the on-site interaction $U$.

\subsection{The strong-coupling limit}\label{subsec:strong}

Given the large number of extended interactions in the model, it is instructive to look at the strong-coupling limit (neglecting all hopping terms) to detect possible charge order instabilities stemming solely from the interactions. 

The reader will forgive us if we use a slightly different notation, writing the interaction Hamiltonian as
\begin{equation}
H_{\rm int} = 
\frac12\sum_{\Rv,\Rv',a,b} V_{\Rv,\Rv'}^{a,b} n_{\Rv}^a n_{\Rv'}^b
\end{equation}
where now $\Rv$, $\Rv'$ denote Bravais lattice sites and $a,b$ orbital indices from 1 to 4.
In essence, for each $\Rv$, the site index $\rv$ takes two values (the two sublattices $A$ and $B$), as does the layer index $\ell$, leading to four possible value of the orbital index $a$.
This shift in notation allows us to express the interaction in Fourier space:
\begin{equation}
H_{\rm int} = \frac12 \sum_{\qv, a,b} \tilde V^{ab}_\qv \tilde n_\qv^{a\dagger}  \tilde n_\qv^b
\end{equation}
where 
\begin{equation}
V_{\Rv\Rv'}^{ab} = \frac1L \sum_\qv \tilde V_\qv^{ab} e^{i\qv\cdot(\Rv-\Rv')} \qquad\qquad 
\tilde n_\qv^a = \frac1{\sqrt{L}}\sum_\Rv e^{-i\qv\cdot\Rv}n_\Rv^a
\end{equation}
Interactions up to third neighbor are then encoded in the following $\qv$-dependent matrix:
\begin{equation}\label{eq:Vq}
[\tilde V_\qv^{ab}] = 
\begin{pmatrix}
U+ V_2\beta_\qv  & V_1\gamma_\qv + V_3 \gamma^*_{2\qv} & V_0 + V_2\beta_\qv  & V_1\gamma_\qv + V_3 \gamma^*_{2\qv} \\
V_1\gamma^*_\qv + V_3 \gamma_{2\qv}  & U+ V_2\beta_\qv & V_1\gamma_\qv + V_3 \gamma^*_{2\qv}  & V_0 + V_2\beta_\qv \\
V_0 + V_2\beta_\qv  & V_1\gamma^*_\qv + V_3 \gamma_{2\qv} & U+ V_2\beta_\qv  & V_1\gamma_\qv + V_3 \gamma^*_{2\qv} \\
V_1\gamma^*_\qv + V_3 \gamma_{2\qv}  & V_0 + V_2\beta_\qv & V_1\gamma^*_\qv + V_3 \gamma_{2\qv}  & U+ V_2\beta_\qv
\end{pmatrix}
\end{equation}
with 
\begin{equation}
\beta_\qv = 2\left(\cos\qv\cdot\bv_1+ \cos\qv\cdot\bv_2+\cos\qv\cdot\bv_3 \right) \quad\mbox{and}\quad
\gamma_\qv = e^{i\qv\cdot\av_1}+ e^{i\qv\cdot\av_2}+ e^{i\qv\cdot\av_3} 
\end{equation}
where the vectors $\bv_i$ are the second-neighbor vectors on the honeycomb lattice (hence first neighbors on the Bravais lattice):
\begin{equation}
	\bv_1 = 2\av_1 + \av_2 \qquad \bv_2 = \av_1 + 2\av_2 \qquad \bv_3 = \av_2-\av_1
\end{equation}
The order of orbitals adopted in this matrix notation is $(w_1,w_4,w_2,w_3)$: the first two orbitals belong to the ``first layer'', the last two to the ``second layer''.

The local density $n_{\Rv\sigma}^a$ can only take the values 0 or 1, but the Fourier transforms $\tilde n^a_\qv$ are continuous variables in the thermodynamic limit, and they all commute with each other.
Hence, for the sake of detecting charge order in the strong-coupling limit, we can treat the variables
$\tilde n^a_\qv$ as classical.

The matrix \eqref{eq:Vq} can be diagonalized  by a unitary matrix: 
\begin{equation}
\tilde V_\qv^{ab} = \sum_{r=1}^4  U^{ar}_\qv \lambda_\qv^{(r)} U^{br*}_\qv 
\end{equation}
and thus the interaction energy can take the form
\begin{equation}
H_{\rm int} = \frac12 \sum_\qv \sum_{r=1}^4 \lambda_\qv^{(r)} |m^{(r)}_\qv|^2  \qquad\qquad
\left( m^{(r)}_\qv  = U^{ar*}_\qv \tilde n_\qv^a \right)
\end{equation}
with the eigenvalues
\begin{align}
&\lambda_\qv^{(1)} = U + V_0 + 2V_2\beta_\qv + 2|V_1\gamma_\qv+V_3\gamma^*_{2\qv}| \\
&\lambda_\qv^{(2)} = U + V_0 + 2V_2\beta_\qv - 2|V_1\gamma_\qv+V_3\gamma^*_{2\qv}| \\
&\lambda_\qv^{(3)} = \lambda_\qv^{(4)} = U -V_0
\end{align}

The uniform solution $\tilde n_{\mathbf0}^a = (1,1,1,1)$ corresponds to $\lambda^{(1)}_{\mathbf0}$, which is the largest possible eigenvalue, and is favored by the (neglected) kinetic energy.
Charge order instabilities in the strong-coupling limit occur for negative eigenvalues, since they can be lower the interaction energy.
When substituting the values given in Eq.~\eqref{eq:values}, one finds that the maximum eigenvalue is $\lambda_{\mathbf 0}^{(1)}=12U$ and the minimum eigenvalue is zero, the latter at the Dirac points $\qv=\Kv$ and $\qv=\Kv'$ for $\lambda^{(1)}_\qv$, and at all wavevectors for $\lambda^{(2,3,4)}_\qv$.
This means that the system has no instabilities in the strong-coupling limit, only indifferent states (zero eigenvalue), especially at wavevectors $\Kv$ and $\Kv'$.
When probing such instabilities with a cluster method, we should therefore make sure that these two wavevectors belong to the reciprocal cluster. 
The 12-site (hexagonal) cluster used in this work statisfies this requirement.
		
\section{The variational cluster approximation}

In order to detect spectral gaps in the normal state and to probe the possible existence of antiferromagnetic or charge-ordered states in this model, we use the variational cluster approximation (VCA)~\cite{potthoff2003,potthoff2003b,potthoff2014a} with an exact diagonalization solver at zero temperature.
This method takes into account short-range correlations exactly, while allowing long-range order through the introduction of broken-symmetry fields determined by a variational principle.

Let us summarize this method, starting with a Hamiltonian containing local interactions only.
We write the lattice Hamiltonian as $H=H_0(\tv)+H_1(U)$, the sum of a noninteracting term $H_0(\tv)$ with one-body Hamiltonian matrix $\tv$, and of an interaction term $H_1(U)$ with a local Hubbard interaction $U$.
If the lattice contains $L$ sites and the model has $B$ orbitals per unit cell, then this matrix $\tv$ is $N\times N$, with $N=LB$.

One then defines a functional $\Omega_\tv[\Sigmav]$ of the self-energy $\Sigmav$ as
\begin{equation}\label{Potthoff1}
\Omega_\tv[\Sigmav]=\Tr\ln\left(-\left(\Gv_0^{-1}-\Sigmav\right)^{-1}\right)
+F[\Sigmav]~~.
\end{equation}
In this expression the trace and the logarithm are functional in nature, $\Gv_0(\omega)=(\omega+\mu-\tv)^{-1}$ is the one-particle Green function of the noninteracting system, and $F[\Sigmav]=\Phi[\Gv[\Sigmav]]-\Tr(\Sigmav \Gv[\Sigmav])$ is the Legendre transform of the Luttinger-Ward functional $\Phi[\Gv]$~\cite{luttinger1960}, $\Gv$ being  viewed a functional of $\Sigmav$. 
The Potthoff variational principle states that $\Omega_\tv[\Sigmav]$ is stationary at the exact, physical self-energy, and its value at that point is the exact thermodynamic grand potential $\Omega$ of the system.

One cannot directly optimize $\Omega$ in Eq.~(\ref{Potthoff1}) since the precise form of $F[\Sigmav]$ is unknown.
But the functional form of $F[\Sigmav]$ depends only on the interaction term $H_1(U)$, not on the one-body term $H_0(\tv)$. 
This motivates us to define a family of simpler, reference Hamiltonians $H'=H_0(\tv')+H_1(U)$ that differ from $H$ in their one-body Hamiltonian matrix $\tv'$ only, for which the Green function $\Gv'(\omega)$, the self-enery $\Sigmav'(\omega)$ and the grand potential $\Omega'$ can be computed numerically. Specifically, $H'$ can be restricted to a small cluster of sites and a numerical method like exact diagonalization can be applied.
Applying Eq.~(\ref{Potthoff1}) to $H'$, we obtain
\begin{equation}\label{Potthoff2}
\Omega_{\tv'}[\Sigmav']=\Omega'=\Tr\ln\left(-\left(\Gv_0^{\prime -1}-\Sigmav'\right)^{-1}\right)+F[\Sigmav'],
\end{equation}
where $\Gv_0'=(\omega+\mu-\tv')^{-1}$ is the noninteracting Green's function for $H'$ and $F$ has the same functional form for both $H$ and $H'$ since they have the same interaction part.
Equation~(\ref{Potthoff2}) then provides an explicit expression for $F$ evaluated at $\Sigma'$:
\begin{equation}
F[\Sigmav']=\Omega'-\Tr\ln\left(-\Gv'\right),
\end{equation}
with $\Gv^{\prime -1}=\Gv_0^{\prime -1}-\Sigmav'$.

So far no approximation was made. The basic approximation of the VCA method is to restrict the space of self-energies $\Sigmav'$
to the physical self-energies of the reference Hamiltonian $H'$ for a suitable set of $\tv'$'s. In other words, we are not making an approximation on the form of the functional $F$, but we restrict the variational space of self-energies: 
We will search a stationary point of $\Omega_\tv[\Sigmav']$ on a subset of one-body terms $\tv'$ in a class of solvable reference Hamiltonians. 
Using Eq.~(\ref{Potthoff1}) and (\ref{Potthoff2}), the functional to be optimized is
\begin{equation}\label{Potthoff3}
\Omega_\tv[\Sigmav'] = \Omega'+\Tr\ln\left(-\left(\Gv_0^{-1}-\Sigmav'\right)^{-1}\right)-\Tr\ln(-\Gv')~~,
\end{equation}
where everything on the r.h.s. can be explicitly computed.

In quantum cluster methods, such as the VCA or cluster dynamical mean field theory, the reference Hamiltonian $H'$ is defined on a set of decoupled (but otherwise identical) \textit{clusters} that tile the lattice exactly.
In other words, $H' = \sum_c H_c$, where $H_c$ is the Hamiltonian for a single cluster containing $N_c$ orbitals, and the sum contains $N/N_c$ terms.
Each cluster must be small enough for $H_c$ to be exactly solvable numerically, say by the Lanczos method or variants thereof.
If the cluster Hamiltonian $H_c$ is simply the restriction of the lattice Hamiltonian to the cluster, i.e., if the variational method described above is not applied, one get the so-called cluster perturbation theory (CPT)~\cite{senechal2000a, gros1993}.
This directly leads to the following approximate Green function
\begin{equation}\label{GCPT}
\Gv^{-1}(\omega)=\Gv_0^{-1}(\omega)-\Sigmav'(\omega)=\Gv^{\prime -1}(\omega)-\Vv,
\end{equation}
where $\Vv=\tv-\tv'$ contains inter-cluster hopping terms that were severed in the reference Hamiltonian and the $N\times N$ matrix $\Sigmav'$ is block diagonal, each block being equal to the self-energy $\Sigmav_c$ of the cluster Hamiltonian $H_c$.

Instead of dealing with $N\times N$ matrices $\Gv$, $\tv$, etc., one can make use of the translation invariance on the superlattice of clusters and express the above relations in terms $N_c\times N_c$ matrices that depend on a wave vector $\kvt$ belonging to the Brillouin zone associated with this superlattice (referred to as the \textit{reduced} Brillouin zone).
The above equation can then be recast as
\begin{equation}\label{GCPT}
\Gv^{-1}(\kvt,\omega)=\Gv_0^{-1}(\kvt,\omega)-\Sigmav_c(\omega)=\Gv_c^{-1}(\omega)-\Vv(\kvt),
\end{equation}
The wave vector $\kvt$ takes $N/N_c$ different values and all quantities of interest are diagonal in this wave vector.
In particular, $\Sigmav_c$ and $\Gv_c$ do not depend on $\kvt$ since all clusters are identical.

If, in the spirit of the Potthoff variational principle, the reference Hamiltonian is not simply the restriction to the cluster of the lattice Hamiltonian but contain additional one-body terms, these will be included in $\Vv(\kvt)$.
Using Eq.~(\ref{GCPT}), the Potthoff functional (\ref{Potthoff3}) will then be written as
\begin{equation}\label{Potthoff4}
\Omega_\tv[\Sigmav']=\Omega'-\Tr\ln\left(1-\Vv\Gv'\right)
\end{equation}
or, in terms of a sum over frequencies and reduced wavevectors,
\begin{equation}\label{Potthoff4}
\Omega_\tv[\Sigmav'] =  \Omega' - \int\frac{d\omega}{2\pi}\sum_\kvt \ln\det\left[\unit-\Vv(\kvt)\Gv_c(\omega)\right]~~,
\end{equation}
where the frequency integral can be taken along the imaginary axis after proper regularization.

In VCA, one searches for stationary points of the functional (\ref{Potthoff4}), i.e., solutions of the Euler equation $\partial\Omega_\tv[\Sigmav']/\partial\tv'=0$. This is achieved in practice by using the cluster one-body terms $\tv'$ as variational parameters. In particular, one can search for spontaneously broken symmetries by including in $\tv'$ symmetry-breaking terms, i.e., Weiss fields. By contrast with conventional mean-field theory, the full dynamical effect of correlations is taken into account via the frequency dependence of the cluster Green's function $\Gv'$ in Eq.~(\ref{Potthoff4}).
In other words, short-range correlations (within the cluster) are treated exactly.

\section{The dynamical Hartree approximation}\label{sec:hartree}

The VCA approximation as summarized above only applies to systems with on-site interactions, since the Hamiltonians $H$ and $H'$ must differ by one-body terms only, i.e., they must have the same interaction part.
This is not true if extended interactions are present, as they are partially truncated when the lattice is tiled into clusters.
To treat the extended Hubbard model, one must apply further approximations.
For instance, we can apply a Hartree (or mean-field) decomposition on the extended interactions that straddle different clusters, while interactions (local or extended) within each cluster are treated exactly.
This is called the dynamical Hartree approximation (DHA) and has been used in Refs~\cite{senechal_resilience_2013, faye2015} in order to assess the effect of extended interactions on strongly-correlated superconductivity.
We will explain this approach in this section.

Let us consider a Hamiltonian of the form 
\begin{equation}\label{eq:Hubbard}
H=H_0(\tv) + \frac12\sum_{i,j}V_{ij} n_i n_j
\end{equation}
where $i,j$ are compound indices for lattice site and orbital, $n_{i\s}$ is the number of electrons of spin $\s$ on site/orbital $i$, and $n_i=n_{i\up}+n_{i\dn}$ (the index $i$ is a composite of honeycomb site $\rv$ and layer $\ell$ indices as used in Sect.~\ref{sec:model}, or of Bravais lattice site $\Rv$ and orbital index $a$ used in Sect.~\ref{subsec:strong}).
The factor $\frac12$ in the last term comes from the independent sums on $i$ and $j$ rather than a sum over pairs $(i,j)$.
In the dynamical Hartree approximation, the extended interactions in the model Hamiltonian \eqref{eq:Hubbard} are replaced by
\begin{equation}\label{eq:hartree}
\frac12\sum_{i,j} V_{ij}^\mathrm{c} n_i n_j  + 
\frac12\sum_{i,j} V_{ij}^\mathrm{ic} (\bar n_i n_j + n_i\bar n_j - \bar n_i \bar n_j)
\end{equation}
where $V_{ij}^\mathrm{c}$ denotes the extended interaction between orbitals belonging to the same cluster, whereas 
$V_{ij}^\mathrm{ic}$ those interactions between orbitals of different clusters.
Here $\bar n_i$ is a mean-field, presumably the average of $n_i$, but not necessarily, as we will see below.
Both the first term ($\hat V^\mathrm{c}$)  and the second term ($\hat V^\mathrm{ic}$), which is a one-body operator, are part of the lattice Hamiltonian $H$ and of the VCA reference Hamiltonian $H'$.

Let us express the index $i$ as a cluster index $c$ and a site-within-cluster index $\a$.
Then Eq.~\eqref{eq:hartree} can be expressed as
\begin{equation}
\frac12\sum_{c,\a,\b} \tilde V_{\a\b}^\mathrm{c} n_{c,\a} n_{c,\b}  + 
\frac12\sum_{c, \a,\b} \tilde V_{\a\b}^\mathrm{ic} (\bar n_\a n_{c,\b} + n_{c,\a}\bar n_\b - \bar n_\a \bar n_\b)
\end{equation}
where we have assumed that the mean fields $\bar n_i$ are the same on all clusters, i.e., they have minimally the periodicity of the superlattice, hence $\bar n_i=\bar n_\a$.
We have consequently replaced the large, $N\times N$ and block-diagonal matrix $V_{ij}^\mathrm{c}$
by a small, $N_c\times N_c$ matrix $\tilde V_{\a\b}^\mathrm{c}$, and we have likewise ``folded'' the large $N\times N$ matrix $V_{ij}^\mathrm{ic}$ into the $N_c\times N_c$ matrix $\tilde V_{\a\b}^\mathrm{ic}$.

In order to make this last point clearer, let us consider the simple example of a one-dimensional lattice with nearest-neighbor interaction $v$, tiled with 3-site clusters.
The interaction Hamiltonian
\begin{equation}
H_{\rm int} = v\sum_{i=0}^N n_i n_{i+1} 
\end{equation}
would lead to the following $3\times3$ interaction matrices:
\begin{equation}
\tilde V^\mathrm{c} = v\begin{pmatrix}0 & 1 & 0\\ 1 & 0 & 1\\ 0 & 1 & 0 \end{pmatrix}
\qquad 
\tilde V^\mathrm{ic} = v\begin{pmatrix}0 & 0 & 1\\ 0 & 0 & 0\\ 1 & 0 & 0 \end{pmatrix}
\end{equation}

In practice, the symmetric matrix $\tilde V^\mathrm{ic}_{\a\b}$ is diagonalized and the mean-field inter-cluster interaction is expressed in terms of eigenoperators $m_\mu$:
\begin{equation}
\hat V^\mathrm{ic} = \sum_\mu D_\mu \left[ \bar m_\mu m_\mu - \frac12\bar m_\mu^2 \right]
\end{equation}
For instance, in the above simple one-dimensional problem, these eigenoperators $m_\mu$ and their corresponding eigenvalues $D_\mu$ are
\begin{align}
D_1 &= -v & m_1 &= (n_1 - n_3)/\sqrt2 \\
D_2 &= \phantom{-}0 & m_2 &= n_2 \\
D_3 &= \phantom{-}v & m_3 &= (n_1 + n_3)/\sqrt2
\end{align}

The mean fields $\bar n_i$ are determined either by applying (i) self-consistency or (ii) a variational method.
In the case of ordinary mean-field theory, in which the mean-field Hamiltonian is entirely free of interactions, these two approaches
are identical. 
In the present case, where the mean-field Hamiltonian also contains interactions treated exactly within a cluster, self-consistency does not necessarily yield the same solution as energy minimization.
In the first case, the assignation $\bar n_i \gets \langle n_i\rangle$ would be used to iteratively improve on the value of $\bar n_i$ until convergence. In the second case, one could treat $\bar n_i$ like any other Weiss field in the VCA approach, except that $\bar n_i$ is not defined only on the cluster, but on the whole lattice. We will follow the latter approach below.

\begin{figure}[htb]\centering
\includegraphics[width=0.5\hsize]{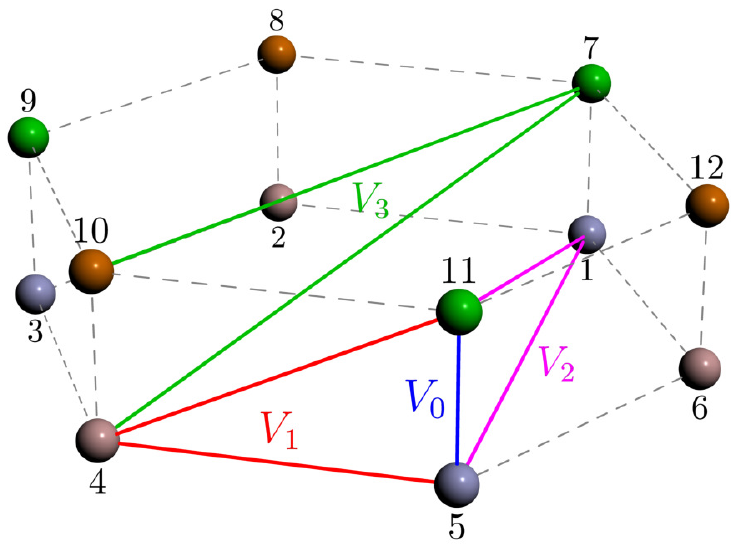}
\caption{12-site cluster used in this work. The extended interactions $V_0$ to $V_3$ are shown.
Different Wannier orbitals are shown as spheres of different colors. Orbitals $w_1$ and $w_4$ are located, say, on the bottom layer, whereas orbitals $w_2$ and $w_3$ are located on the top layer.
}
\label{fig:cluster}
\end{figure}
\setcounter{MaxMatrixCols}{12}
\begin{table}\label{table:Vic}
\[\tilde V^\mathrm{ic} =
\begin{pmatrix}
0 & V_3 & 2V_2 & V_1 & 2V_2 & V_3 & 0 & V_3 & 2V_2 & V_1 & 2V_2 & V_3 \\
V_3 & 0 & V_3 & 2V_2 & V_1 & 2V_2 & V_3 & 0 & V_3 & 2V_2 & V_1 & 2V_2 \\
2V_2 & V_3 & 0 & V_3 & 2V_2 & V_1 & 2V_2 & V_3 & 0 & V_3 & 2V_2 & V_1 \\
V_1 & 2V_2 & V_3 & 0 & V_3 & 2V_2 & V_1 & 2V_2 & V_3 & 0 & V_3 & 2V_2 \\
2V_2 & V_1 & 2V_2 & V_3 & 0 & V_3 & 2V_2 & V_1 & 2V_2 & V_3 & 0 & V_3 \\
V_3 & 2V_2 & V_1 & 2V_2 & V_3 & 0 & V_3 & 2V_2 & V_1 & 2V_2 & V_3 & 0 \\
0 & V_3 & 2V_2 & V_1 & 2V_2 & V_3 & 0 & V_3 & 2V_2 & V_1 & 2V_2 & V_3 \\
V_3 & 0 & V_3 & 2V_2 & V_1 & 2V_2 & V_3 & 0 & V_3 & 2V_2 & V_1 & 2V_2 \\
2V_2 & V_3 & 0 & V_3 & 2V_2 & V_1 & 2V_2 & V_3 & 0 & V_3 & 2V_2 & V_1 \\
V_1 & 2V_2 & V_3 & 0 & V_3 & 2V_2 & V_1 & 2V_2 & V_3 & 0 & V_3 & 2V_2 \\
2V_2 & V_1 & 2V_2 & V_3 & 0 & V_3 & 2V_2 & V_1 & 2V_2 & V_3 & 0 & V_3 \\
0 & 2V_2 & V_1 & 2V_2 & V_3 & 0 & V_3 & 2V_2 & V_1 & 2V_2 & V_3 & 0
\end{pmatrix}\]
%
\begin{center}
\includegraphics[width=1.0\hsize]{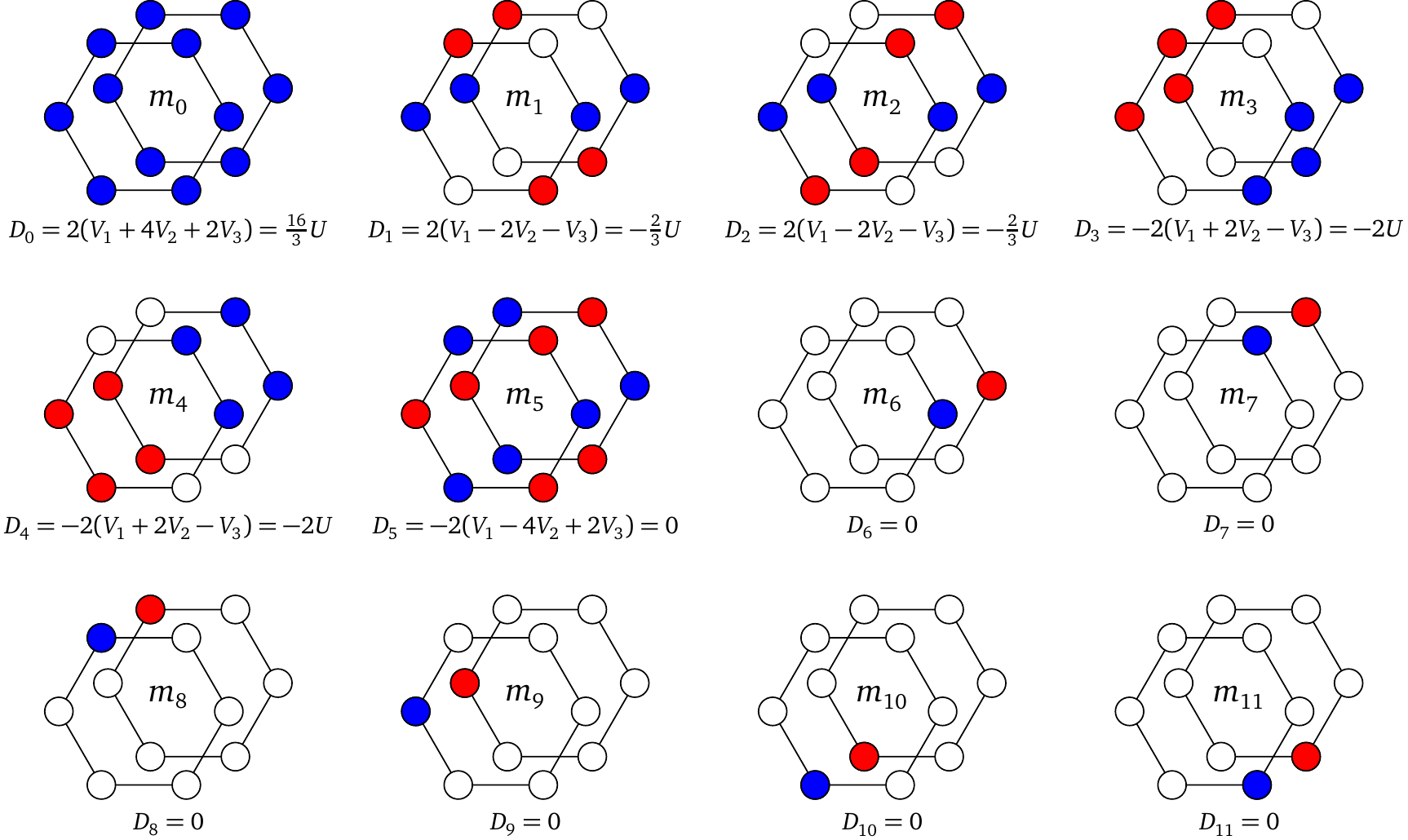}
\end{center}
\caption{Inter-cluster coupling matrix for the 12-site cluster used in this work.
The numbering of sites is illustrated on Fig.~\ref{fig:cluster}. Bottom: eigenvalues $D_\mu$ and corresponding eigenvectors (or eigenoperators) $m_\mu$ of this matrix. The eigenoperators are shown graphically as a function of site on the 12-site cluster: blue means 1 and red $-1$.
The eigenvalues are also shown as a function of the on-site repulsion $U$ when the constraints~\eqref{eq:values} are applied.
\label{table:Vic}}
\end{table}

\begin{figure}[htbp]\centering
\includegraphics[width=\hsize]{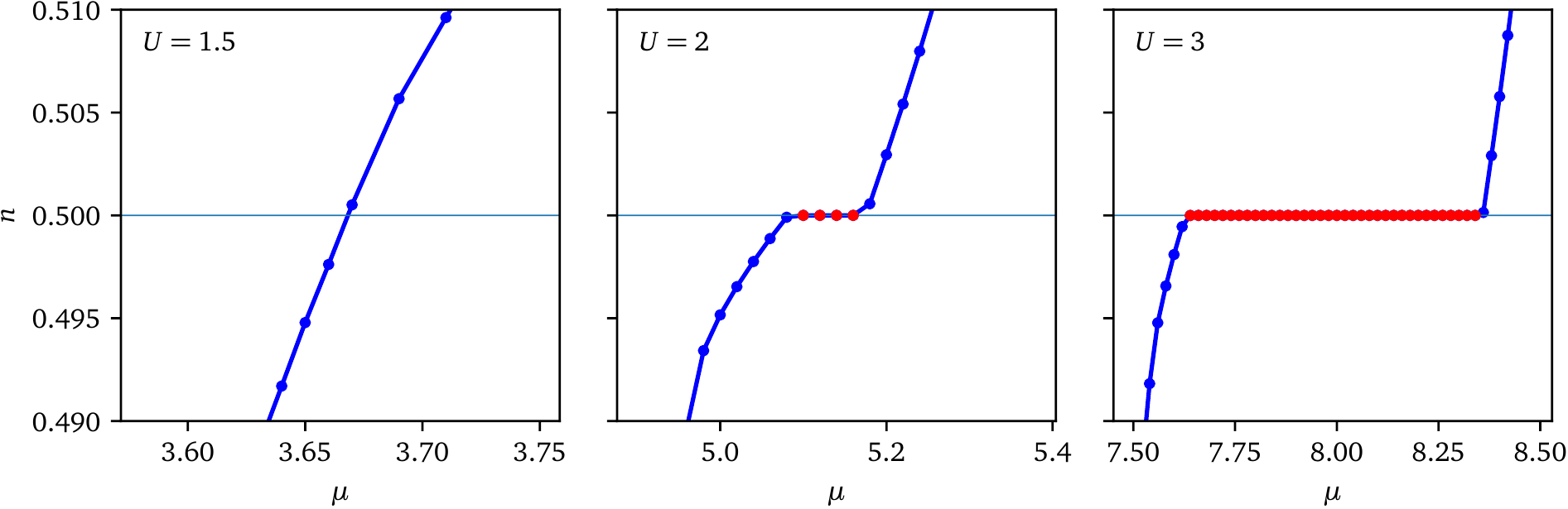}
\caption{Electronic density vs chemical potential $\mu$ for different interaction strengths at quarter filling.
The presence of a plateau (in red) is the signature of an insulating state, and the width of the plateau is the magnitude of the gap.
The insulator-to-metal transition occurs between $U=1.5$ and $U=2$.
}
\label{fig:sec6_density}
\end{figure}

\begin{figure}[htbp]\centering
\includegraphics[width=\hsize]{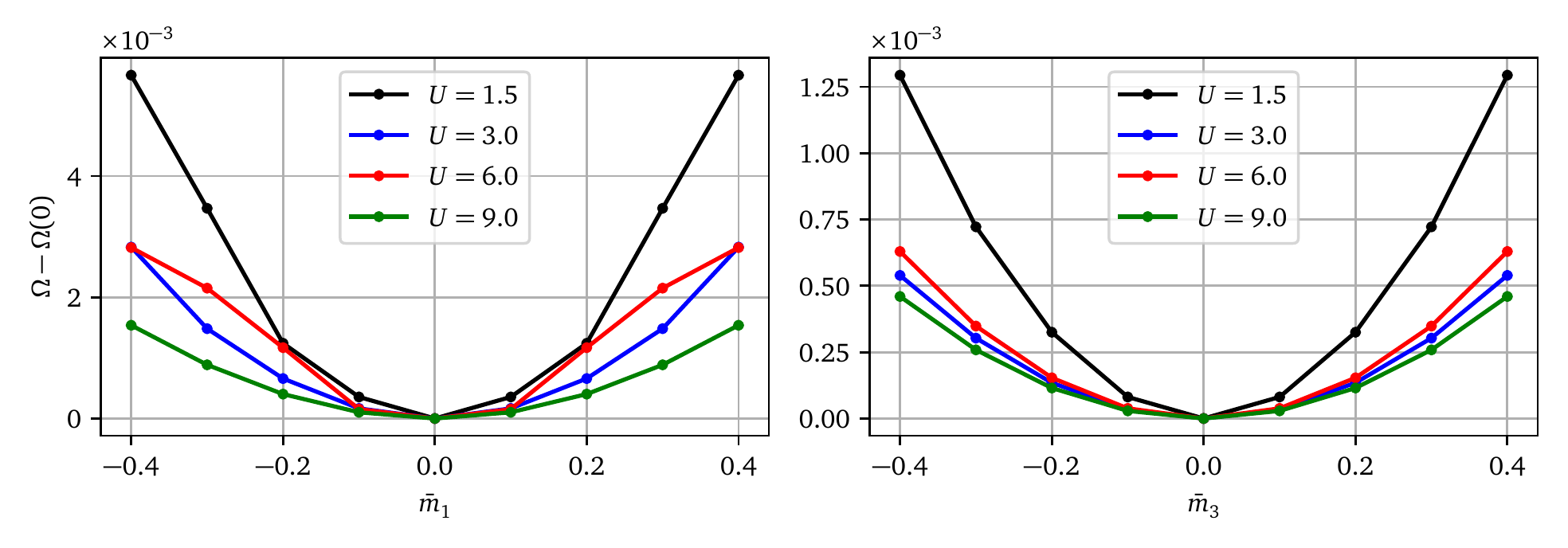}
\caption{Left panel : The Potthoff functional $\Omega$ as a function of the charge-density-wave Weiss field $\bar m_1$ at quarter-filling. Right panel: the same, for the charge-density-wave Weiss field $\bar m_3$.
See Table~\ref{table:Vic} for an illustration of the density-waves $m_1$ and $m_3$.
The symmetric state (no charge density wave) $\bar m_{1,3}=0$ is stable.
}
\label{fig:omega_cdw3}
\end{figure}

\section{The normal state at quarter filling}

In this work we use a 12-site cluster containing 3 unit cells of the low-energy model. 
It is made of two superimposed hexagonal clusters, as illustrated on Fig.~\ref{fig:cluster}.
On that figure the various extended interactions $V_0$ to $V_3$ are indicated.
The three wavevectors of the reciprocal cluster are $\mathbf\Gamma=\mathbf0$, $\Kv$ and $\Kv'$.
The $12\times12$ matrix of inter-cluster interactions is given in Table~\ref{table:Vic} and the eigenoperators $m_\mu$ used in the dynamical Hartree approximation are illustrated in the lower part of the same table.

We begin by investigating the normal state at quarter filling, for several values of the interaction $U$, all the extended interactions following from $U$ according to Eq.~\eqref{eq:values}.
We will start by applying VCA to detect the insulating state, assuming that no charge order is present.
To do this, we treat the cluster chemical potential, $\mu_c$, as a the sole variational parameter in the VCA procedure.
We do not take into account inter-cluster interactions, i.e., the Hartree approximation described in Sect.~\ref{sec:hartree}.
Indeed, all the sites of the 12-site cluster are equivalent in the absence of charge order, meaning that the relevant (normalized)  eigenvector of the inter-cluster interaction matrix $V^\mathrm{ic}$ is 
\begin{equation}
m_0 = \frac1{2\sqrt3}\sum_{i=1}^{12}n_i 
\end{equation}
Therefore, adding the corresponding mean-field $\bar m_0 m_0$ to the lattice Hamiltonian would simply shift the chemical potential by $-\bar m_0$, and leave the variational space used in VCA unchanged.
This would therefore not help us in determining whether there is a gap or not.

The signature of the Mott gap will be a plateau in the relation between $\mu$ and the density $n$.
This is shown in Fig.~\ref{fig:sec6_density} for a few values of the interaction $U$. 
Using the cluster chemical potential $\mu_c$ as a variational parameter makes the plateaux very sharp, whereas not using VCA, i.e., simple cluster perturbation theory (CPT) would make the plateaux softer, thereby making the transition to the metallic state more difficult to detect.
In the case shown, the metal-insulator transition clear occurs between $U=1.5$~meV and $U=2$~meV.
This Mott transition is essentially caused by extended interactions.

The question then arises as to the nature of the insulating state at quarter filling: is there a charge density wave or not?
As shown in Sect.~\ref{subsec:strong}, the charge fluctuations are expected to be large, because a full array of charge configurations do not affect the energy in the strong-coupling limit when the extended interactions follow Eq.~\eqref{eq:values}.
We do expect, on intuitive grounds, that the kinetic energy terms would be unfavorable to charge order.
Nevertheless, in order to probe the possible existence of charge order, we will apply Hartree inter-cluster mean-field theory, as described in Sect.~\ref{sec:hartree}. In order to put all the chances on our side, we will probe one of the eigenoperators with the lowest (negative) eigenvalues in Table~\ref{table:Vic}, namely one of those with $D=-2$:
\begin{equation}
m_{3} = \frac1{2\sqrt2}\left(n_1 + n_2 - n_4 - n_5 + n_7 + n_9 - n_{10} - n_{11} \right)
\end{equation}
We must then optimize the Potthoff functional as a function of the mean field $\bar m_3$, in addition to using $\mu_c$ as a variational parameter. On the right panel of Fig.~\ref{fig:omega_cdw3} we show the Potthoff functional $\Omega$ as a function of $\bar m_3$ for a value $\mu_c$ that actually optimize $\Omega$ at a value of $\mu$ associated with quarter filling, for a few values of the interaction $U$. This is to illustrate the absence of nontrivial solution for $\bar m_3$, i.e., the value of the mean-field parameter $\bar m_3$ that minimizes the energy is indeed zero. 
This shows that, within this inter-cluster mean-field approximation and for these values of $U$, there is no charge order this type ($m_3$ or, equivalently, $m_4$) at quarter-filling.

We perform the same computation for the $m_{1}$ eigenoperator:
\begin{equation}
m_{1} = \frac1{2\sqrt2}\left(n_1 - n_2 + n_4 - n_5 + n_7 - n_9 + n_{10} - n_{11} \right)
\end{equation}
and find similar results, as shown on the left panel of Fig.~\ref{fig:omega_cdw3}.
Therefore, for the values of $U$ probed, the quarter-filled state appears to be a pure, uniform Mott insulator, driven by extended interactions.

\begin{figure}[htbp]\centering
	\includegraphics[width=1.0\hsize]{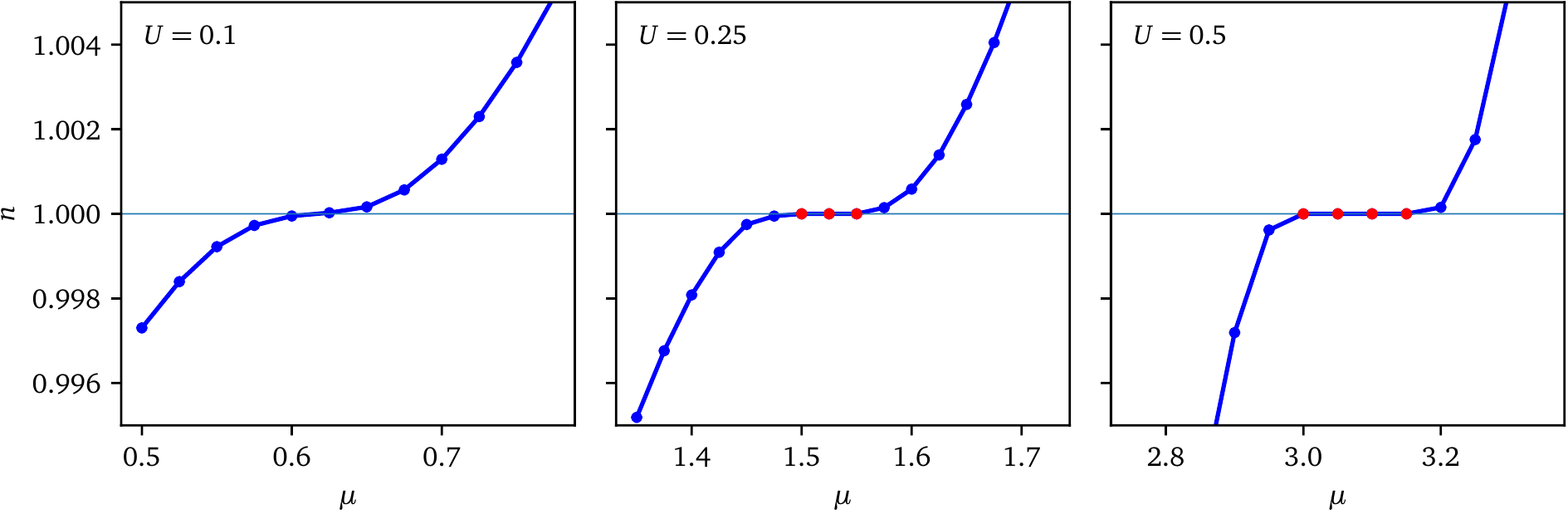}
\caption{Electronic density vs chemical potential $\mu$ for different interaction strengths at half filling, similar
to Fig.~\ref{fig:sec6_density}.
The presence of a plateau (in red) is the signature of an insulating state, and the width of the plateau is the magnitude of the gap.
The insulator-to-metal transition occurs between $U=0.1$~meV and $U=0.25$~meV.
}
	\label{fig:sec12_density}
\end{figure}

\begin{figure}[htbp]\centering
	\includegraphics[scale=0.9]{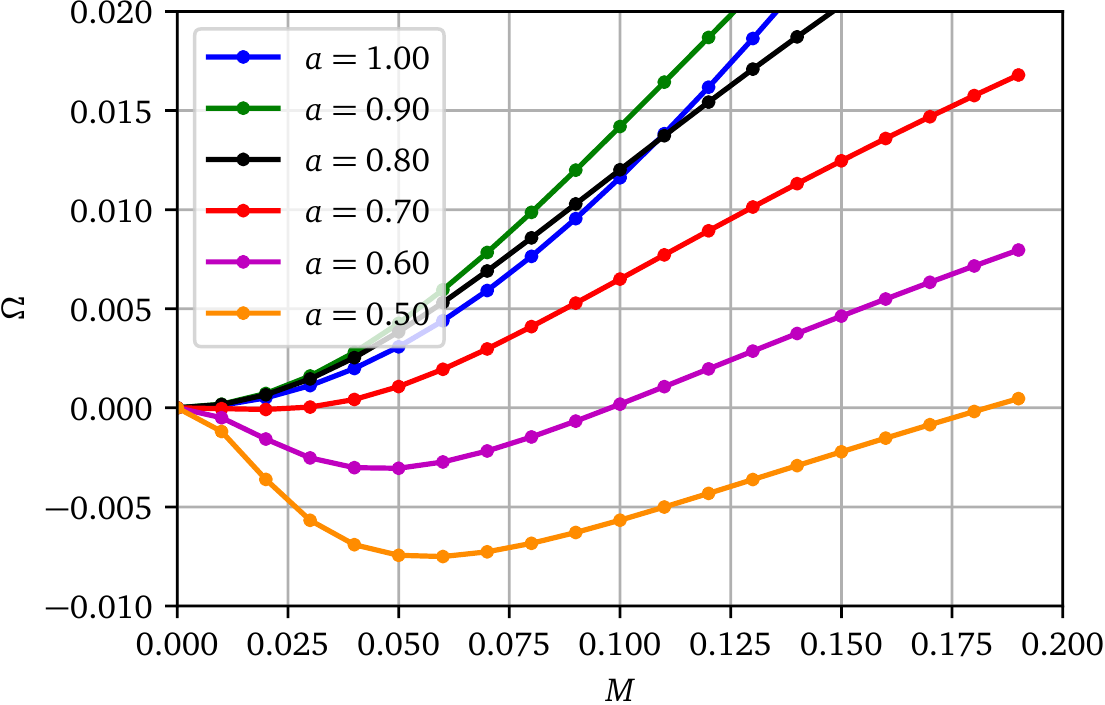}
\caption{Potthoff functional vs the antiferromagnetic Weiss field $M'$ for several values of $a=3V_1/2U$ and $U=3$, at half-filling.
The case $a=1$ corresponds to the contraints~\eqref{eq:values}, and smaller values of $a$ just weaken the extended interactions compared to the on-site interaction. The value of $\Omega$ at $M=0$ is subtracted for clarity.
Antiferromagnetism appears only below $a=0.7$, i.e., not for the extended interactions constrained by Eq.~\eqref{eq:values}.
}
\label{fig:omega_M}
\end{figure}

\section{The normal state at half filling and antiferromagnetism}

The insulating state at half-filling is revealed the same way as at quarter-filling, by applying the VCA with $\mu_c$ as a variational parameter.
The results are shown in Fig.~\ref{fig:sec12_density}, where it appears that the Mott transition occurs between $U=0.1$~meV and $U=0.25$~meV, i.e., at a much lower value of the interaction than at quarter filling.

We will not probe charge order at half-filling, as an antiferromagnetic state is more expected to occur.
The Weiss field used to probe antiferromagnetism is
\begin{equation}
\hat M = M\sum_{i=1}^{12} (-1)^i (n_{i\uparrow}-n_{i\downarrow})
\end{equation}
Fig.~\ref{fig:omega_M} shows the Potthoff functional as a function of $M$ for different values of the extended interactions compared to the on-site repulsion $U=3$~meV.
These different values are characterized by the ratio $a=3V_1/2U$, which is unity when the extended interactions obey the constraints \eqref{eq:values}.
Otherwise, the extended interactions $V_{0-3}$ have the same ratios between them as in Eq.~\eqref{eq:values}.
Lower values of $a$ correspond to weaker extended interactions (compared to $U$).
From that figure we see that, even at a relatively strong $U$ (the Mott transition occurs at a much lower value of $U$), antiferromagnetism is not present at half-filling for the nominal values of the extended interactions defined in Eq.~\eqref{eq:values}. Upon lowering these interactions, antiferromagnetism appears.
Hence the half-filled state should be a true Mott insulator, not an antiferromagnetic insulator.

This is relatively easy to understand in the strong-coupling limit, when Eq.~\eqref{eq:values} holds. The low-energy manifold at half-filling in the absence of hopping terms is degenerate non only because of spin, but also because of charge motion: if there is exactly one electron on each site, hopping an electron to the neighboring site does not change the interaction energy, and thus the usual strong-coupling perturbation theory argument leading to an effective Heisenberg model at half-filling and large $U$ does not hold anymore. 

\section{Conclusion}

We have probed the insulating states at quarter- and half-filling in a tight-binding model for magic angle twisted bilayer graphene, augmented with local and extended density-density interactions.
For a wide range of interactions obeying the constraints \eqref{eq:values}, we have detected the Mott gap using the variational cluster approximation (VCA) with a 12-site cluster and located the Mott transition between $U=1.5$~meV and $U=2$~meV at quarter filling, and between $U=0.1$~meV and $0.25$~meV at half-filling.
In addition, we have investigated the possibility of charge order at quarter-filling using the VCA and an inter-cluster Hartree approximation for the extended interactions, and concluded that it does not arise.
Lastly, we have probed antiferromagnetism at half-filling and concluded likewise that it does not arise when the extended interactions obey the relations \eqref{eq:values}.
It looks therefore plausible that the correlated insulating states observed at these filling ratios are genuine Mott insulators and not gapped ordered states.

\paragraph{Funding information}
DS acknowledges support by the Natural Sciences and Engineering Research Council of Canada (NSERC) under grant RGPIN-2020-05060. Computational resources were provided by Compute Canada and Calcul Québec.


\end{document}